\documentclass[10pt]{iopart}
\expandafter\let\csname equation*\endcsname=\relax
\expandafter\let\csname endequation*\endcsname=\relax

\usepackage{amsmath}
\usepackage{subcaption}
\usepackage{graphicx}
\usepackage{citesort}
\usepackage{xcolor}
\usepackage{amsmath}

\usepackage{iopams}  
\usepackage{tabularx}
\newcolumntype{Y}{>{\centering\arraybackslash}X}
\newcommand{\phyp}{\textit{Physarum polycephalum}~}

\graphicspath{{./images/}}

\begin{document}

\title{Network emergence and reorganization in confined slime moulds}
	
\author{
Raphaël Saiseau$^{1}$, Valentin Busson$^{1}$, Laura Xénard$^{1}$, Marc Durand$^{1}$}

\address{$^{1}$Laboratoire MSC, Universit\'{e} Paris Cité, CNRS, UMR 7057, Mati\`{e}re et Syst\`{e}mes Complexes (MSC), F-75006 Paris, France.}

\ead{marc.durand@u-paris.fr}
\vspace{10pt}
\begin{indented}
	\item[]December 2022
\end{indented}




\begin{abstract}
A fundamental question regarding biological transport networks is the interplay between the network development or reorganization and the flows it carries.  
 We use \emph{Physarum polycephalum}, a true slime mould with a transport network which adapts quickly to change of external conditions, as a biological model to make progress in this question. We explore the network emergence and reorganization in specimens suddenly confined in chambers with ring geometry. Using an image analysis method based on the structure tensor, we quantify the emergence and directionality of the network. We show that confinement induces a reorganization of the network with a typical $10^{4}$s timescale, during which veins align orthoradially along the ring. We show that this network evolution relies on local dynamics. 
\end{abstract}

\vspace{2pc}

%
%
\maketitle
%
\ioptwocol


	
\section{Introduction}
Biological transport networks are crucial for the functioning of many organisms. Examples include the vasculature of vertebrates (from aorta to capillary bed) and plants (from roots to leaf venation), the bronchial system, and the mycelium of fungal colonies. Such self-organized structures exhibit a number of properties (efficiency, adaptability, resilience) that are highly desirable for technical applications. The understanding of their formation and evolution has also obvious medical applications, as numerous diseases are associated with pathological evolutions of their structure. 
It is well documented that growth and remodelling of biological transport networks are influenced by mechanical factors \cite{Nguyen_2006,Alkilani_2008}.



A  challenging issue regarding biological transport networks is to understand the interplay between flow and network emergence and reshaping (coined respectively as vasculogenesis and angiogenesis for the vascular network of vertebrates). The slime mould \emph{Physarum polycephalum} in its plasmodium stage,  is a giant multinucleated single cell organism that develops a network of tubular elements in which flows are generated by contractions of the actomyosin cortex contained in the surrounding membrane~\cite{alim2013random,alim2017mechanism}.  
Even in the absence of a pacemaker like the heart, cortex dynamics can self-organize to give rise to coordinated flows on large scales \cite{kuroda2015allometry,takagi2008emergence,takagi2010annihilation,busson2022emergence,zhang2017self}. In comparison, the interplay between this contractile activity, the flow they induce, and the network self-organizing dynamics has been poorly investigated: usually, network architecture is considered as static elements when studying contractile activity~\cite{julien2018oscillatory,radszuweit2014active,oettmeier2019lumped}.  Actually, segmenting the network structure and tracking its evolution are heavy computational tasks.

In spite of its apparent simplicity, the development of this network shares common features with the development of vascular systems in higher organisms \cite{risau_1997}, or with the mechanisms that take place in the irrigation of tumours \cite{Pries_2009}. 
In particular, one can clearly identify two stages in the development of the network: a growing phase during which \emph{P. polycephalum} develops a dense and reticulated network of small tubular elements. Then a reorganizing phase during which the network becomes more hierarchical and less reticulated \cite{Tero_2010}. The self-organizing behaviour of its tubular network has shown to be useful to solve complex problems~\cite{Fricker2009adaptive}. 
Its two-dimensional growth and rapid change in external conditions make it a model organism to identify the mechanisms involved in the formation and evolution of biological transport networks.
In a previous study \cite{busson2022emergence}, we investigated the stable contractile patterns emerging in a physarum plasmodium confined in an annular chamber. This geometry has numerous advantages: it reduces the problem to a quasi-unidimensional system while preserving the structural heterogeneities of a (macro)plasmodium.
Moreover, the periodic boundary conditions suppress antero-posterior axis and so pre-established polarity of
the giant cell, while confinement and deprivation of nutrients prevent plasmodium growth or displacement. 

In the present paper, we investigate the emergence and reshaping of the transport network for plasmodia confined in annular geometries. For this purpose, rather than using 
computationally heavy network segmentation tools, we use a coarse-grained description based on the structure tensor \cite{Jahne_1993,puspoki_2016}.

\begin{figure}
	\centering
		\includegraphics[width=\columnwidth]{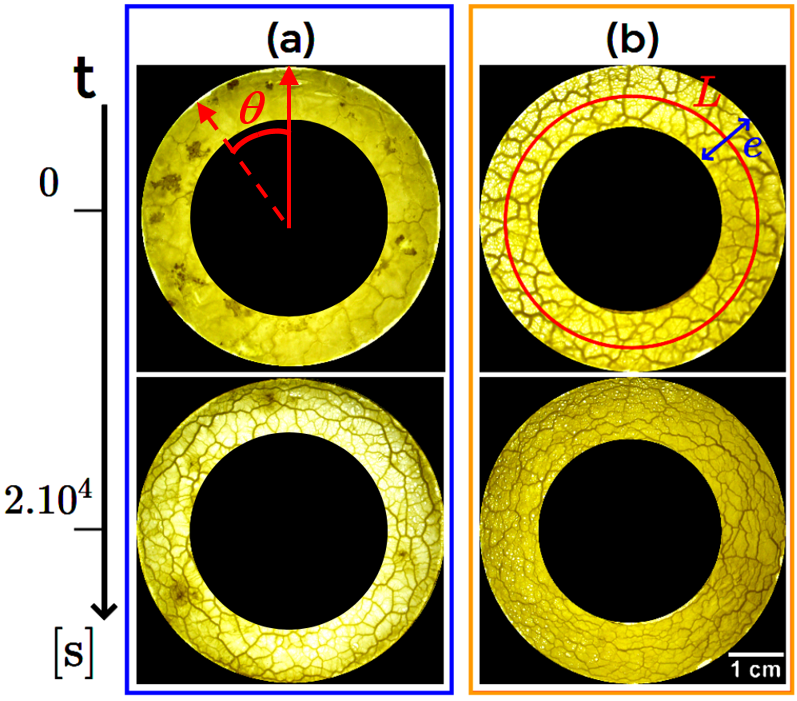}
	\captionof{figure}{Bright field microscopy images showing (a) network emergence and (b) network reorganization for ring confined Physarum. Raw images are taken at $t=0~$s and $t=20000~$s.}
	\label{fig:Reorganization_image}
\end{figure}

\section{Methods \label{sec:Methods}}

The experimental set-up we use is the same that has been used to observe contractile oscillations patterns \cite{busson2022emergence}: \phyp plasmodia are confined in ring-shaped chambers of controlled geometrical dimensions. The ring perimeter $L$ (measured at the center) varies from 6.0 to 13.5 cm whereas the aspect ratio $L/e$, with $e$ the width of the ring, varies from 11 to 41. These high aspect ratios allow us to consider a pseudo 1-D description of the specimen. To make ring shaped plasmodia, the plasmodium is first inseminated from previous cultivation on a Petri dish half filled with aqueous gel. Multiple insemination points are disposed on a gel in an initial large disk. When the gel is covered by the plasmodium homogeneously, ring plastic walls are used to cut multiple ring shaped specimen concentrically. Two different initial configurations occur, and are analyzed separately in Section \ref{sec:results} (see Fig. \ref{fig:Reorganization_image}): either the ring-shaped plasmodium has no visible tubular network, and then we analyze how annular confinement affects its emergence. Or an isotropic network is already present in the plasmodium, and then we analyze how the confinement affects its reshaping. The Petri dish is then closed and sealed, and after a typical settling time of 15–30 min, the video recording starts using transmitted light microscopy. RGB images are taken every 4 or 6 seconds for films typically lasting 4 to 12 hours (see Fig.\ref{fig:Reorganization_image}). 

\begin{figure}
	\centering
	\includegraphics[width=0.9\columnwidth]{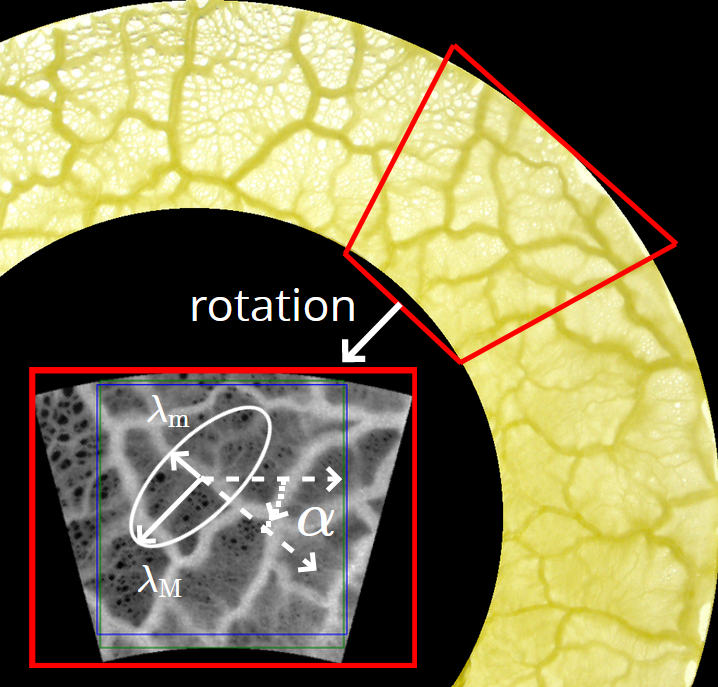}
	\captionof{figure}{Schematics of the successive operations made on a given angular sector: angular sector rotation, blue channel selection and structure tensor derivation. $\lambda_{\mathrm{M}}$, $\lambda_{\mathrm{m}}$ are the eigenvalues of the structure tensor with $\lambda_{\mathrm{M}} > \lambda_{\mathrm{m}}$ and $\alpha$ is the angle between the ellipse big axis direction and the local horizontal coordinate.}
	\label{fig:Angular_sectors}
\end{figure}

From the movies, we use an image analysis method to obtain an exhaustive characterization of network emergence and reorganization. Since the specimen is mostly yellow, we minimize the computational time while maximizing the image contrast by using only the blue channel of the RGB images. 
As Figures \ref{fig:Reorganization_image} and \ref{fig:Angular_sectors} reveal, the network architecture is difficult to segment from the plasmodium because of the large range of spatial scales it spans and the inhomogeneity of plasmodium thickness. Instead of using computationally costly and perfectible segmentation tools, we use coarse-grained gradient-based orientation estimators measured at the meso-scale (the local network architecture) to characterize it.


To quantify emergence and orientationality or the network, we calculate the structure tensor \cite{Jahne_1993,puspoki_2016}, defined on every point $\mathbf{x_0}$ as
\begin{equation}
	\mathbf{J(\mathbf{x_0})} =  
	\begin{pmatrix}
		\overline {I_x^2} &	\overline {I_x I_y} \\
		\overline {I_x I_y} & \overline {I_y^2}
	\end{pmatrix}
\end{equation}
 where $I_{x}=\partial_x I$ and $I_{y}=\partial_y I$ are the partial derivatives of the intensity $I$ I with respect to $x$ and $y$. The weighted averaging is defined as 
 \begin{equation}
 	\overline {f g }(\mathbf{x_0}) = \int_{\mathbb{R}^{2}} f(\mathbf{x}) g(\mathbf{x} ) w_R (\mathbf{x}-\mathbf{x_0}) d \mathbf{x}
 \end{equation} where $w_R$ is a Gaussian window of radius $R$
that defines the meso-scale size over which the estimators are defined. 
 $\mathbf{J}$ is a $2 \times 2$ symmetric positive-definite matrix, and we note $\lambda_{\mathrm{M}}$, $\lambda_{\mathrm{m}}$ ( $\lambda_{\mathrm{M}}> \lambda_{\mathrm{m}}>0$) its two positive eigenvalues, defined for each position $\mathbf{x_0}$.  

We then define the local energy $E$,  the local coherency $C$ and the local gradient orientation $\psi$ as:
\begin{align}
		E(\mathbf{x_0})= \lambda_{\mathrm{M}} + \lambda_{\mathrm{m}},\\
 	C(\mathbf{x_0})= \dfrac{\lambda_{\mathrm{M}} - \lambda_{\mathrm{m}}}{\lambda_{\mathrm{M}} + \lambda_{\mathrm{m}}} ,\\
 	\tan \left(2	\psi(\mathbf{x_0})\right)=   \dfrac{2\overline{ I_x I_y} }{\overline{I_x^2}- \overline{I_y^2}}.
\end{align}
The so-called energy $E$ quantifies the magnitude of the local gradient, and then is related with the emergence of veins. The coherency $C$ is a measure of the local gradient anisotropy, and indicates if the local image features are oriented or not: $C=1$ when the local structure has one dominant orientation and $C=0$ if the image is essentially isotropic in the local neighbourhood. The angle $\psi \in [-\pi/2,+\pi/2]$ corresponds to the orientation of the eigenvalue associated with $\lambda_M$ and then indicates the main local gradient orientation. Accounting for the symmetry with respect to the radial direction, we are mainly interested in its norm $\vert \psi \vert$. Note that the main local vein orientation is orthogonal to it.

To obtain tractable datasets, we decompose the ring in small angular sectors of uniform arc length $ s \simeq 3.3~$mm. The radius of the Gaussian weighting function is set to $R =\sqrt{  s^{2} + e^{2}} $, and the above quantities are derived within the maximum rectangle inscribed in the angular sector to avoid border artefacts in the gradient evaluations . Values of $\vert \psi \vert$, $C$ and $E$ are then averaged over this rectangular area. Moreover, each sector is rotated as shown in Fig.\ref{fig:Angular_sectors} before calculating $\vert \psi \vert$ and its average value $\langle \vert \psi \vert \rangle$ so that angles are defined with respect to the local orthoradial direction of the ring. We finally define the mean vein orientation in the sector, with respect to its orthoradial axis, as $\alpha=\pi/2-\langle \vert \psi \vert \rangle$ ($\alpha \in [0,\pi/2]$).
An orientation $\alpha = 0$ (respectively $\alpha =\pi /2$) corresponds to veins mainly aligned along the orthoradial (respectively radial) direction.
Given this window size, the spatial scale over which these estimators are defined is given by the angular sector size independently of its aspect ratio.
We have tested with regular structures of lines having radial and orthoradial orientations and with modulated intensity that $E$, $C$ and $\alpha$ evolve as expected (see S.I.).



\section{Results \label{sec:results}}

Figures. \ref{fig:Reorganization_image}(a) and \ref{fig:Reorganization_image}(b) illustrate the two kinds of experimental initial conditions, respectively a homogeneous plasmodium without network, and a plasmodium containing an isotropic network. We analyze separately the network evolution starting from these two initial configurations. Moreover to test the effect of nutrients availability  \cite{dussutour2010amoeboid}, two gel compositions have been used: one composed of $2\%$ Phytagel and $1\%$ glucose only, and the other of $2\%$ Oatmeal agar, $1\%$ Phytagel and $1\%$ glucose. 
The presence of Oatmeal agar clearly favours the presence of homogeneous plasmodium without network \cite{takamatsu_2009}. 

\subsection{Network emergence in confined homogeneous plasmodium \label{chap:emergence}}


To quantify the emergence of the network in the initially homogeneous plasmodium, we consider the coherency weighted energy $E_{\mathrm{C}} = CE/\sum_{\theta} C$. This weighting favours energy variation only directly related to the emergence of veins over other phenomena inducing intensity variations.
\begin{figure*}[htb]
	\centering
	\includegraphics[width=\textwidth]{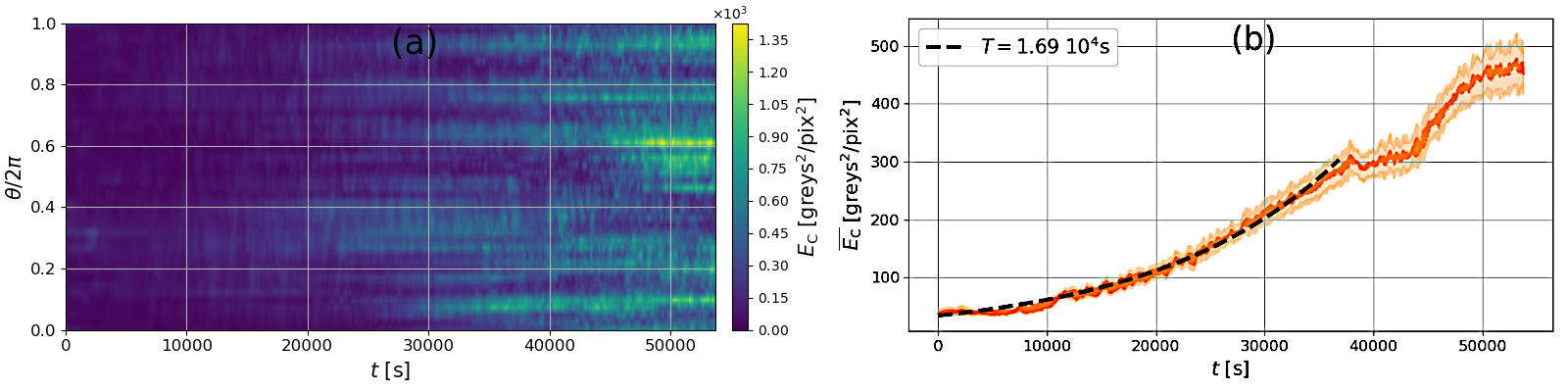}
	\includegraphics[width=\textwidth]{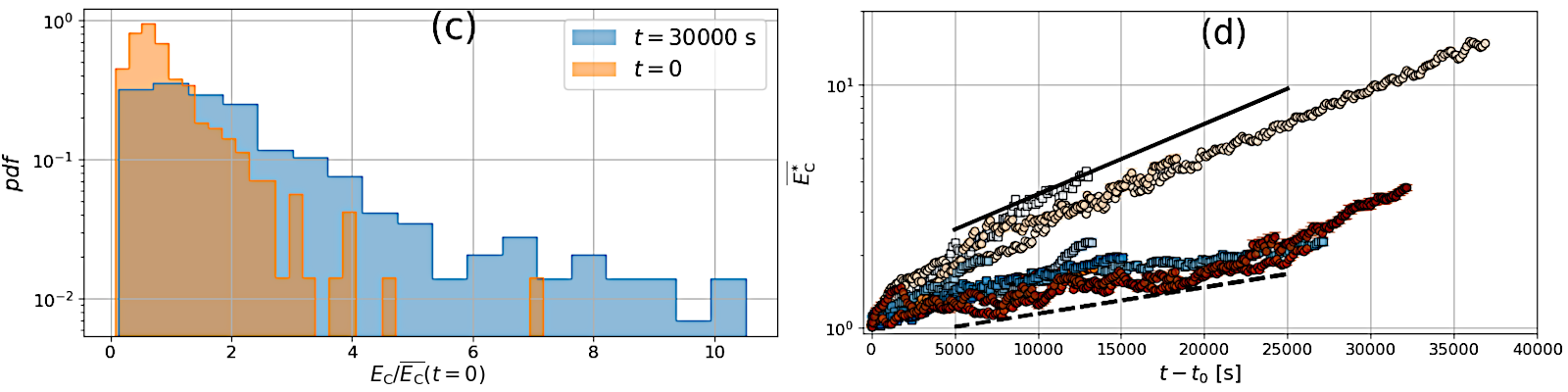}
	\caption{(a) Space-time plot of weighted energy $E_{\mathrm{C}}$ for the specimen shown in Fig.\ref{fig:Reorganization_image}(a). (b) Same data averaged over $\theta$. Dashed black curve: exponential fit $e^{t/T}$ with adjusted parameter $T=1.69\times 10^4$s. Shaded area indicates standard error of the mean. (c) Histograms of the weighted energy distribution $E_{\mathrm{C}}/\overline{E_{\mathrm{C}}}(t=0)$ at times $t=0$ and $t=30000~$s calculated over the 11 samples. The ratio of both mean values is $\overline{E_{\mathrm{C}}}(t=30000~\text{s})/\overline{E_{\mathrm{C}}}(t=0) = 2.53$. (d) Coherency weighted energy scaled by its initial value and averaged over $\theta$
		$\overline{E^{*}_{\mathrm{C}}} = \langle E_{\mathrm{C}} (\theta,t) / E_{\mathrm{C}} (\theta,t=0) \rangle_{\theta} $	for all experiments starting with an homogeneous plasmodium without network. Orange rounded markers (from light to dark) correspond to specimens for which the gel was initially supplied with Oatmeal, whereas blue squared markers (from light to dark) correspond to specimens deprived of Oatmeal. Error bars indicating the standard error of the mean are smaller than or close to the marker size. Two exponential curves $e^{t/T_\mathrm{E}}$ are shown respectively in solid black line with timescale $T_\mathrm{E}=1.5 \times 10^4$s, and in dashed black line with $T_\mathrm{E}=4.0 \times 10^4$s. Mean timescale value is $T_\mathrm{E}=1.95 \pm 0.86 \times 10^4$s with the uncertainty given by the standard deviation of fitted timescale distributions. 
}
	\label{fig:Energy_emergence}
\end{figure*}

Figure \ref{fig:Energy_emergence}(a) shows the space-time plot of $E_{\mathrm{C}}$ for the plasmodium shown in Fig.\ref{fig:Reorganization_image}(a). We observe a global increase of the weighted energy for all the angular positions $\theta$, revealing the emergence of a network. It can be noticed that the pattern shows important variations between the different angular positions, suggesting that the network emergence is the result of local dynamics.

The increase in weighted energy is made more visible when plotting its value averaged over $\theta$, as shown in Fig.\ref{fig:Energy_emergence}(b). On a large part of the plot, we observe an exponential behavior with a time scale $T=1.69\times 10^4$s.   Note that at the end of the experiments, blebbing and network pruning (associated with optimization of transport, see \cite{marbach2016pruning}) are causing departure from this exponential growth. 
To avoid any external effects in the dynamics characterization, we now only consider part of the experiments where the emergence dynamics is clearly visible, \emph{i.e.} where no pruning, blebbing or vein meandering occur (see S.I.). We call $t_{0}$ the initial time of these time windows.

Figures \ref{fig:Energy_emergence}(c) and \ref{fig:Energy_emergence}(d) show the data for 11 different confined plasmodia without initial network (for 8 initially different organisms). Figure \ref{fig:Energy_emergence}(c) shows the probability distribution of the coherency weighted energy $E_\mathrm{C}$ scaled by its initial $\theta-$averaged value $\overline{E_{\mathrm{C}}}(t=0)$, at the initial time and $3\times10^4\mathrm{s}$ after. We clearly observe a large spreading of the histogram towards high $E_\mathrm{C}$ values, corresponding to the emergence of a network in those plasmodia.

\begin{figure*}[ht]
	\centering
	\includegraphics[width=\textwidth]{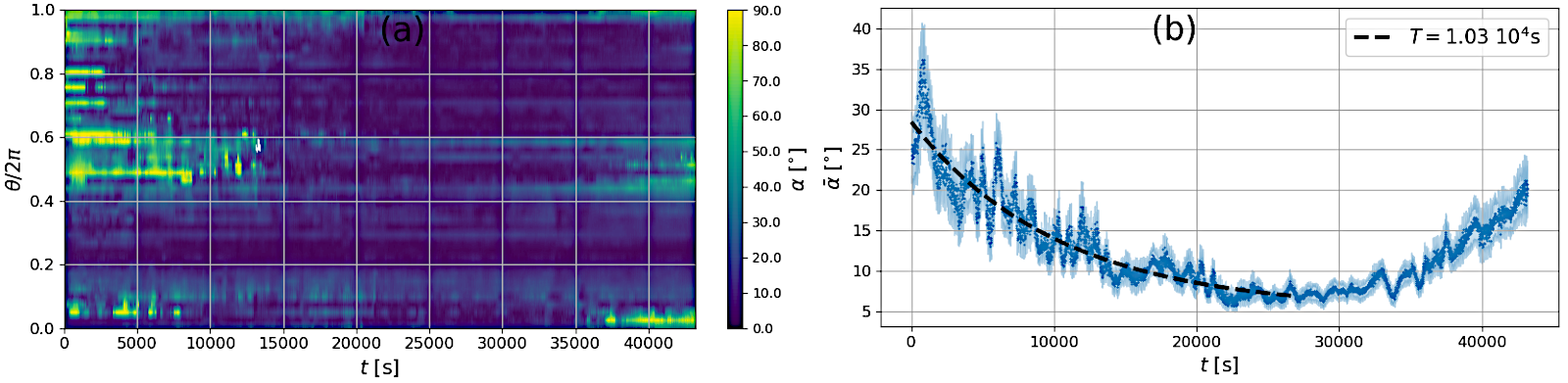}
	\includegraphics[width=\textwidth]{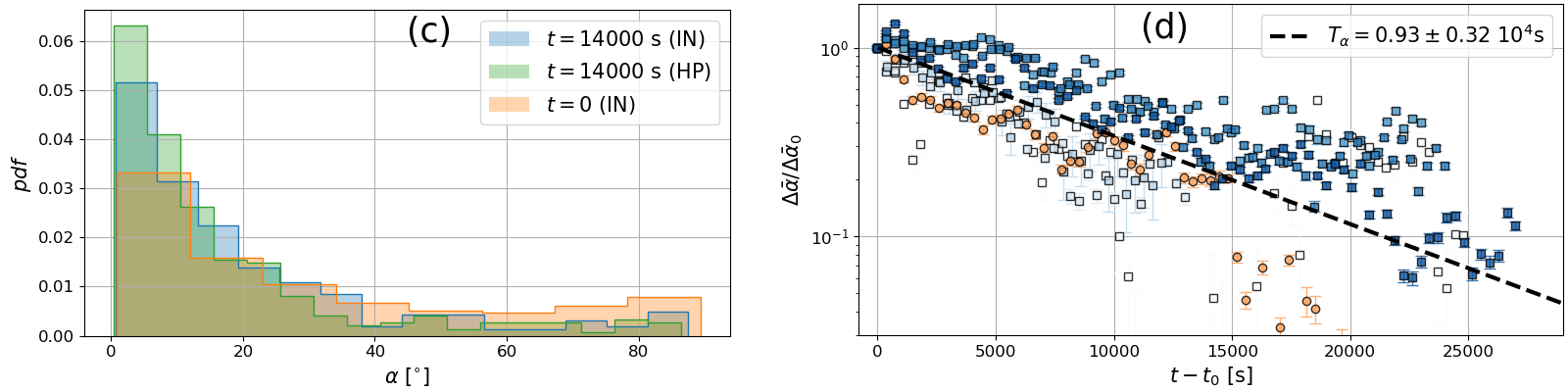}
	\caption{(a) Space-time plot of network directionality $\alpha$ for the specimen shown in Fig.\ref{fig:Reorganization_image}(b). (b) Same data averaged over $\theta$. Shaded area indicates standard error of the mean. Dashed black line: exponential fit $e^{-t/T}$ with adjusted parameter $T=1.03~10^4$s. (c) Histograms of $\alpha$ values at times $t=0~$ and $t=14000~$s, calculated for every annular sector and over the 7 specimens with isotropic network initial condition (IN), and at $t=14000~$s calculated for every annular sector and over the 11 specimens with homogeneous plasmodium initial condition (HP). The mean values are $\bar{\alpha} (t=0, $IN$) = 30 ^{\circ}$, $\bar{\alpha}(t=14000~$s$, $IN$) = 21 ^{\circ}$ and $\bar{\alpha}(t=14000~$s$, $HP$) = 17 ^{\circ}$. (d) Rescaled mean network direction $\Delta \bar{\alpha} / \Delta \bar{\alpha}_{0}$ over time with $\Delta \bar{\alpha}_{0} = \Delta \bar{\alpha} (t=0)$ for the 7 specimens with isotropic network initial condition (IN). Orange rounded markers (from light to dark) correspond 
		to specimens in which the gel was initially supplied with Oatmeal, whereas blue squared markers (from light to dark) correspond to specimens deprived of Oatmeal. Dashed black line: exponential fit $e^{-t/T_\mathrm{\alpha}}$ with $T_\mathrm{\alpha}=(0.93 \pm 0.32)\times 10^4$s being the unique free parameter adjusted over all data. Error bars indicate the standard error of the mean.}
	\label{fig:Directionality_reorganization}
\end{figure*}

Figure \ref{fig:Energy_emergence}(d) shows the time evolution of  $\overline{E^{*}_{\mathrm{C}}} = \langle E_{\mathrm{C}} (\theta,t) / E_{\mathrm{C}} (\theta,t=0) \rangle_{\theta} $, the coherency weighted energy scaled by its initial value and averaged over $\theta$ for the 11 samples.
The data can all be fitted using exponential law $\exp({t/T_\mathrm{E}})$ with timescales $T_\mathrm{E}$ ranging between $1.0 \times 10^4 $ and $4.0 \times 10^4$s. The mean value is estimated to be $T_\mathrm{E}=(1.95 \pm 0.86) \times 10^4$s, with the uncertainty given by the standard deviation of the $T_\mathrm{E}$ distribution. The variability in timescale is not correlated to variability in ring geometry, nutrients availability, or any clear visible morphological features apart from a seemingly darker color. This color may indicate some internal effect (cell life cycle, organism aging or water content).


Let us emphasize that the network emerges with veins oriented preferentially along the orthoradial direction (see Fig. \ref{fig:Reorganization_image}(a)). This will be discussed in more details in the next section by comparing it with the network reorganization from isotropic network initial conditions.

\subsection{Reshaping of confined isotropic network}

We now consider plasmodia that contain an isotropic spanning network at the time they are punched in the annular chambers. We use the $\alpha$ angle introduced in Sect. \ref{sec:Methods} to quantify the local network directionality relatively to the orthoradial direction and characterize the network reorganization.
Figure \ref{fig:Directionality_reorganization}(a) shows the space-time plot of $\alpha$ for the specimen shown in Fig.\ref{fig:Reorganization_image}(b). We observe an overall decrease of $\alpha$ values over time (except close to the end of the experiments) indicating a reshaping of the network during which veins are aligning with the orthoradial direction of the ring. The slight increase of $\alpha$ at the end of the experiment corresponds to the appearance of blebs and especially vein meandering. The artifical increase in  $\alpha$ due to vein meandering is investigated more deeply in the S.I. 
As with the network emergence studied in \ref{chap:emergence}, the signal heterogeneity along the ring observed in Fig.\ref{fig:Directionality_reorganization}(a) suggests that the network reshaping is the result of local dynamics. 

The value of $\alpha$ averaged over the ring angular position $\theta$, plotted in Fig. \ref{fig:Directionality_reorganization}(b), shows that the dynamics associated with network reorganization is well adjusted by an exponential decaying function with typical timescale $T=1.03\times 10^{4}$s.

In order to focus on reorganization dynamics, we now only consider time windows where network spatial reorganization is clearly visible and where blebbing and meandering do not occur. As before, we note $t_{0}$ the initial time in such a time window.
Figures \ref{fig:Directionality_reorganization}(c) and \ref{fig:Directionality_reorganization}(d) show the data for 7 different confined plasmodia having an initial isotropic network (for 6 initially different specimens).
 Figure \ref{fig:Directionality_reorganization}(c) shows  the probability distribution of the network directionality $\alpha$ at times $t=0~$s and $t=14000~$s (before vein meandering). The distribution at the latter time is clearly more peaked around $\alpha=0^\circ$, showing that the confinement induces an alignment of the veins with the orthoradial direction of the ring.
 We also reported in Fig. \ref{fig:Directionality_reorganization}(c) the probability distribution of $\alpha$ measured over the 11 specimens with no initial network, at time $t=14000~$s. 
 The distribution is also peaked around $\alpha=0^\circ$ showing that for a confined plasmodium, the veins emerge with an orientation preferentially aligned with the ring main axis. The corresponding mean $\alpha$ value is $17^\circ$, slightly below the mean value ($21^\circ$) for specimens with initial isotropic network, suggesting that $\alpha$ has not reached its stationary value for the reorganizing case.
  
For all specimens having an initial isotropic network, the evolution of the $\theta-$averaged value of directionality $\bar\alpha$ follows an exponential decay and can be fitted with
\begin{equation}
	t\rightarrow (\alpha_0-\alpha_\infty)e^{-t/T}+\alpha_\infty
	\label{eq:exp_fit}
\end{equation}
The free parameters $\alpha_0$, $T$ and $\alpha_\infty$ represent respectively the initial mean orientation, the decay time, and the final mean orientation in absence of artefacts (blebs, meandering).
Figure \ref{fig:Directionality_reorganization}(d) shows that $\Delta \bar{\alpha}= \bar{\alpha} - \alpha_\infty$ follows an exponential behaviour with a common time scale $T_{\alpha}=0.93 \pm 0.32~10^4$s which corresponds to the mean value of the fitted time scales $T$. The error bars correspond to the standard deviation of the distribution of $T$. Besides veins meandering and blebs appearance, network reticulation and pruning also occur on long time scales. The second phase observed on some experiments where the directionality begins to increase again show these effects and more generally the difficulty to obtain a completely unambiguous behaviour.

Interestingly, the timescale $T_{\alpha}$ of network reorganization is close to the timescale $T_{E}$ that characterizes the network emergence, suggesting that both dynamics are driven by a common mechanism. As before, the kinetics of network reorientation does not seem to be affected by the initial presence or not of proteins in the gel. 

\section{Discussion and conclusion}

Using an image analysis method relying on the structure tensor, we have shown that the confinement of \emph{Physarum polycephalum} plasmodia in chambers with annular geometry affects the architecture of its network: when the network is not present at the initial time, it emerges with veins preferentially aligned with the orthoradial direction. When an isotropic network exists before confinement, it reorganizes with veins aligning orthoradially. The typical timescale observed for network emergence or reorganization is a few $10^{4}~$s, two orders of magnitude slower than the period of contractile oscillations \cite{busson2022emergence}. Our results indicate that in the interplay between contractile wave orientation and network orientation, the former clearly precedes the latter: geometrical confinement orients contractile patterns that in turns orient cytoplasmic flow.  The network is then shaped by the integrated flows over time~\cite{guy2011flow}.

Note that a $10^{4}$s timescale has also been reported by Rodiek and Hauser \cite{rodiek2015migratory} as the typical time for freely migrating microplasmodia to attain diffusive regime. Below this timescale, migration follows ballistic motion. Hence, the typical timescale for diffusive regime may be limited by the time required for the network to reorganize and adapt to a new contractile wave orientation.
Note that although our image analysis based on the structure tensor has shown to be well-suited to detect the emergence and alignment of the veins with the orthoradial direction of the annular chamber, it is limited on long times by the meandering of veins. Although vein meandering has already been reported in other studies \cite{takamatsu_2009}, its origin remains elusive so far.

\ack
We acknowledge financial support from French National Research Agency Grants ANR-17-CE02-0019-01-SMARTCELL and CNRS MITI ‘Mission pour les initiatives transverses et interdisciplinaires’ (reference: BioRes).

\section*{References}

\providecommand{\newblock}{}


\end{document}